\documentclass[12pt]{article}
\usepackage{graphicx}

\def\bfDelta{\mathord{\mbox{\boldmath $\Delta$}}}

\begin{document}
\begin{center}
{\bf MULTISCALE BEHAVIOR AND FRACTIONAL KINETICS  FROM THE DATA OF  SOLAR WIND - MAGNETOSPHERE COUPLING}
\end{center}

\begin{center}
{\bf
G.M. Zaslavsky$^{(1,2)}$, P. N. Guzdar$^{(3)}$, M. Edelman$^{(1)}$, M. I. Sitnov$^{(3)}$ and A. S. Sharma$^{(4)}$
\\
$^{(1)}$Courant Institute of Mathematical Sciences,
New York University, 

251 Mercer St., New York, New York 10012, USA \\
$^{(2)}$Department of Physics, New York University,
2-4 Washington Place, New York, New York 10003, USA \\
$^{(3)}$Institute for Research in Electronics and Applied Physics,

University of Maryland, 
College Park, Maryland 20742, USA \\

$^{(4)}$Department of Astronomy,  
University of Maryland, 

College Park, Maryland 20742, USA 
}

{\it 09 November 2005\\ mscale-pop.tex}
\end{center}

\newpage
\noindent
{\bf ABSTRACT}

Multiscale phenomena are ubiquitous in nature as well in 
laboratories.   A broad range of interacting space and time
scales determines the dynamics of many systems which are
inherently multiscale. In most research disciplines multiscale phenomena
are not only prominent, but also they have often played the dominant role.  
In the solar wind - magnetosphere interaction, multiscale features coexist
along with the global or coherent features.  Underlying these
phenomena are the mathematical and theoretical approaches such as
phase transitions, turbulence, self-organization, fractional kinetics,
percolation, etc. The fractional kinetic equations provide a suitable 
mathematical framework for multiscale behavior.  
In the fractional kinetic equations the multiscale nature is described 
through fractional derivatives and the solutions of these equations yield
non-convergent moments, showing strong multiscale behavior.  
Using a L\'{e}vy-flights approach, we analyze the data of 
the magnetosphere and the solar wind. Based on this
analysis we propose a model of the multiscale features and
compare it with the solutions of
diffusion type equations. The equation with fractional spatial
derivative shows strong multiscale behavior with divergent moments. 
On the other hand the equation with space dependent diffusion
coefficients yield convergent moments, indicating Gaussian type 
solutions and absence of long tails typically associated with 
multiscale behavior.

\newpage
\noindent
{\bf 1. INTRODUCTION}

Many laboratory and natural systems have   a broad range of interacting
space and time scales and their  dynamics exhibit  multiscale
features. The time-series data 
of such systems  have some common characteristics:  the physical variables 
reveal  the properties often referred to as   multiscale phenomena, heavy tails, strong intermittency, etc. Examples of such systems include  dynamical systems
with chaos, stock market indices, earthquakes, space and laboratory plasmas, atmospheric and hydrological systems, etc. 
A variety of characteristic quantities, such as  Hurst
exponents, multifractal spectra, Levy flights, etc. have been computed
from the time series data  using different approaches and
techniques.   Typically, the application of a specific method  has been largely motivated by the current understanding of the relevant processes in a particular system, and their relationship to the available data.  

 From the dynamical systems point of
view, the multiscale properties suggest  fractal or multifractal structure of the dynamical processes
and numerous simulations support this idea [1]. 
The advantage of the analysis of data generated by chaotic dynamics is a
possibility to link the  system evolution to the equations
of the processes and to their  origins, e. g.,  physical, chemical  or
biological 
processes. For example, the L\'{e}vy flights can be identified as 
dynamical processes of trapping into potential wells or ballistic
propagation. Observing and analyzing different realizations of stochastic
dynamical processes, one can classify them and apply the available 
techniques to the phenomena of which the details  of the
processes  are not sufficiently known.
A technique similar  to the one used in the analysis of L\'{e}vy flights has
been developed 
recently and has been applied to the plasma density fluctuations
in the 
tokamak device [2]. Many   studies of  the solar
wind - magnetosphere coupling 
have used the time series data to obtain the Hurst exponent
characterizing the multiscale 
nature [3,4,5]. These studies have motivated 
the modeling of  multiscale behavior in terms of well-known equations 
such as Fokker-Planck   [6, 7] and Ginzburg-Landau 
equations  [8]. 

It is however very desirable to apply different techniques to the same
data set to obtain a consistent characterization of the underlying
processes, and thus provide a strong basis for developing models.   
This point is underscored by the fact that the knowledge of the Hurst
exponent  [9, 10]  does not provide  sufficient
information about the physical processes underlying the data. Indeed, the expression,   $<[x(t_1)-x(t_2)]^2> \sim |t_2-t_1|^{2H}$,
which defines the Hurst exponent $H$, in fact, does not provide  a clear and
unique information about the random process $x(t)$. 
This is an important issue and three pertinent   points are elaborated in
the following.  

First,  the data typically characterize some specific physical observable
  $x(t)$ and the Hurst exponent is strictly related to that
  observable. Other physical variables will have different exponents
  and the connection between these exponents is not known in the absence of    more information about the real processes. Practically, it means that
the common characterization as   ``subdiffusion'' for $H<1/2$ and ``superdiffusion'' for
$H>1/2$ have no real meaning unless  the physical
processes underlying  the diffusion or, more generally, the kinetic processes, are specified.   Second, the Hurst exponent is related to the second
moment  only, and does not reflect the behavior of other moments.   
  For example, for Gaussian or Poissonian processes  
   the second moment is sufficient to characterize the system as all higher moments can be
  expressed in terms of it.   Some Gaussian type processes can have
  $H<1/2$, and the higher moments 
can be evaluated from the data
  that yield  the Hurst exponent. However, this feature can be revealed only  by computing the higher moments. For the  L\'{e}vy and other multiscale processes the second and higher  moments are  infinite.   Such features are not captured by the Hurst exponent even if  $H>1/2$
since, as mentioned above,   the process may be
superdiffusive, but it can not lead to a definite  identification.   Third, the Hurst exponent represents the time dependence of the variance but the
 time characteristic of the process,
typically, is not sufficient, as is clear in the case of  chaotic dynamics [1].  Indeed, the physical process can be characterized by a probability
distribution function $F(x,t)$ and the moments of $x$ are defined as 
$<x^m(t)>=\int x^m(t) F(x(t),t)dx(t).$ In order to distinguish, for example,  a multiscale process
from a Gaussian one,  more subtle  features from the
data than just the Hurst exponent are needed. It should be noted that even for a purely
diffusive process for which $F(x,t)$ satisfies a kinetic equation with derivatives
with respect to $x$ and $t$, not only the exponent $H$ is
important for the kinetics but also the deviation $\delta x$ during some
specified interval  $\delta t$ is required.

The important role of kinetics  was effectively demonstrated in [zaslavsky00, 2] where a new technique of data analysis was applied to the data of plasma density
fluctuations in a tokamak. Assuming the presence of a L\'{e}vy-type
multiscale  process characterized by ``flights'', 
 a special procedure  was used to separate the flight events
without a characteristic  scale from the  ``noisy'' events with a
characteristic small scale. The notion of ``flight'' is however non-trivial
and it should be defined carefully  to avoid ambiguity. It should be pointed out that recent studies of the data of space plasmas [6, 7] did not define the   
L\'{e}vy flights in their analysis.  This issue deserves more careful attention [1]) and will be discussed in more detail in the following.  

A simple way of  introducing the notion of a L\'{e}vy-type flight, or in 
short, flight, is to consider a function $x(t)$ or its integral(s) and
obtain the long   intervals  $\delta t$ where the changes in  $x(t)$
or its integral(s) are monotonic up to some level of accuracy. The change
$\delta x$ in  $x(t)$ is then a measure of the length of 
a flight with the distribution function 
$ F(\delta x, \delta t)$, within the specified level of accuracy.  It is very important to note that   $ F(\delta x, \delta t)$ is the distribution of
flights if they exist! Further,   all moments 
$<x^m(t)>$, obtained from the data, are always finite, i.e.,  $ F(\delta x,
\delta t)$ is truncated and  
defined up to  $\delta x_{max}$, $\delta t_{max}$. Then the
extrapolation of   $ F(\delta x, \delta t)$ should be carried out with utmost care.   For example, in the case of 
 $ F(\delta x, \delta t) \sim 1/(\delta x)^k$,  the moments $<( \delta x)^q>$
with $q \ge k-1$ will be infinite, but   the observational data
can yield   the exponent $k$ within the interval 
$0 \le \delta x \le  \delta x_{max}$.

In this paper  a specific form of L\'{e}vy flight analysis [2]  to the data  of the solar wind-magnetosphere
coupling. A correlated dataset [11] of the 
magnetospheric dynamics driven by the solar wind is used in this study. In this dataset corresponding to the last peak of the solar cycle,  the solar wind (SW) is represented by the flow-induced electric field and the magnetospheric response is represented by the auroral electrojet index AL. 
This data set is used to study   the multiscale
properties by applying a technique based on an a priori assumption of the existence of Levy-like flights. 
This analysis yields a new characterization of the magnetospheric dynamics. 

Multiscale behavior has been modeled using the  so-called fractional generalization of the Fokker-Planck equation [1].  
In order to determine the multiscale properties  of such equations,
numerical solutions of specific cases of such equations are studied. In
the case of scale dependent diffusion coefficients it is found that the
solutions of the equations yield   convergent moments, and thus are not
suitable for modeling the multiscale properties. In the case when the
equations have fractional derivatives, viz. they represent fractional
kinetics, and the numerical solutions do not yield convergent moments. 
It should be noted that the diffusion equation, 
even with a fractional time derivative representing the so-called
fractal Brownian motion, has 
solutions in the form of a Gaussian function $G(x/t^a),$  for which all the
moments are finite for any exponent $a$.  
In the case of a kinetic equation with a fractional derivative 
with respect to coordinate or momentum
one can get the same exponent but 
only few (or even a single) first moments are finite, and all higher moments
 are infinite  for any $t$. 
  In the case when  the kinetic process  $ F(x, t)$ is modeled using
the Fokker-Planck type equation with a scale dependent diffusion
coefficients, as  in [7], we show that all moments are finite.
 This leads to the conclusion that a correct exponent for the second moment 
alone is not sufficient to construct a model, and it is essential that the 
model yields the appropriate multiscale behavior.
 
The structure of the tails of the distributions of different physical
observables becomes crucial  (see also [12]),   and as mentioned above, for data of finite length, defining the tail is a fairly sensitive task
and needs confirmation from independent measurements. 
In particular,  a specific demonstration is provided on how the accuracy of the
data for the tail can be improved (see Fig.~4) if the relevant processes
exhibit flights.

The structure of the paper is as follows.  Section 2 presents a brief 
overview of the solar wind - magnetosphere interaction, the database and 
the recent studies relevant to this paper. 
In Sec. 3  the  L\'{e}vy flight technique used to 
analyze the data [2] is described and also
presents the corresponding results. The analysis indicate a power-law type behavior of the distribution functions  in time and in the magnitudes of the fluctuations.  The multiscale character of the observational data and some mathematical
models for their description are discussed in Sec.4, followed by a detailed analysis of the
fractional kinetic equation in Sec. 5. The reconstruction of the  kinetic equations from the data is discussed in Sec. 6. Numerical
studies of the fractional kinetic equation is described in Sec. 7, and  the conclusions of the paper are summarized in Sec. 8.\\

\noindent
{\bf 2. COUPLED SOLAR WIND - MAGNETOSPHERE SYSTEM}
 
The Earth's magnetosphere is a huge cavity, mostly shielded from the flow of 
charged particles coming from the Sun, the   solar wind plasma,   by its 
magnetic field. The 
solar wind plasma can enter this magnetic shield through  
magnetic reconnection. The main source of plasma for the 
magnetosphere is the ionosphere, a relatively thin and dense plasma shell, 
which separates the magnetosphere from the Earth's thermosphere and 
atmosphere. The solar wind flow is not steady, because of the active solar 
corona, e. g., solar flares, coronal mass ejections and other bursty processes, 
as well as   the inherent turbulent nature of the solar wind flow 
towards the Earth. Similarly, the magnetospheric activity is determined 
both by its interactions with the  solar wind and by the complex internal activations on 
many scales occurring in the practically collisionless plasma of the 
magnetosphere. Attempts to classify the variety of observed phenomena  have 
resulted in descriptive classifications  such as  magnetospheric storms, 
substorms, pseudo-breakups, convection bays, saw-tooth events, etc.  Yet, the description  remain essentially incomplete. 
Moreover, with some exceptions, e.g., the storm-substorm relationship, they do not reflect important connections and similarities 
between processes occurring on different spatial and temporal scales, 
which are revealed both in the solar wind [13, 14] and in the magnetosphere [15, 16].

The recent studies of the  multiscale behavior in the coupled solar wind -
magnetosphere system have been based on two approaches. In the first, the
observational data from different sources, such as spacecraft-borne
and ground-based instruments, have been used to analyze the scaling
properties. In the second approach, mathematical models such as
Fokker-Planck type equations have been used to interpret these
processes. In the latter equations which are known to exhibit
multiscale properties are used to study the conditions under which the
scaling laws and exponents derived from data can be obtained. 
However it should be noted that 
the ability of a particular type of equation to fit the data does not
exclude other types of equations from reproducing the desired scaling,
etc. Such studies provide valuable results that may lead to the  origins
of the multiscale behavior.  These two approaches, viz. the studies using
data and the modeling using equations, are discussed in the following 
sections. 
 
The magnetospheric activity is characterized by a set of parameters, 
including both the basic physical ones, such as the magnetic and electric 
fields, plasma density etc., and the specific geophysical parameters or 
indices, e. g., Kp, Dst, AL, AU, AE, etc. [17]. A particularly relevant set of indices is the family of 
Auroral Electrojet 
indices AE, AU and AL,  obtained from a number 
(usually greater than 10) of stations distributed in local time in the 
latitude region that is typical of the northern hemispheric auroral zone. 
For each of the stations the north-south magnetic perturbation H is 
recorded as a function of the universal time. A superposition of these data 
from all the stations yields a lower bound or maximum negative excursion 
of the horizontal component of the magnetic field variations, and is called the AL index. An upper 
bound, defined similarly, is called the AU index and the difference between these two indices, 
AU-AL, is called the AE index. Auroral indices are particularly 
relevant for substorm studies and have been used  extensively
for more general nonlinear time series analysis and forecasting 
[18, 19] and for multiscale 
studies [16].

The scaling properties inherent in a data set can be represented by many 
characteristic parameters. One of the most popular is the Hurst exponent H 
[9, 10] measuring the "roughness" of multifractal data and allowing one to distinguish between  conventional diffusion (H=0.5), subdiffusion 
($H<0.5$) and superdiffusion ($H>0.5$), as discussed earlier in Section 1.  
However, the presently available 
results for the Hurst exponent,  computed largely from the AE index are rather 
controversial, can not distinguish between sub- and 
super-diffusion regimes. In particular, Takalo and Timmonen [3] found 
H=0.5 based on the AE index studies. Uritsky and Pudovkin [20] concluded 
that H is largely less than 0.5 (H=0.38-0.59, depending on the substorm 
phase). In contrast, Price and Newmann [21] found that H is mostly greater   than 0.5 (H=0.44-0.97, depending on the time scale range). In a 
series of studies involving extensive data processing,  Hnat et 
al. [4, 5, 22] used   different geophysical variables,  including the  solar wind parameter $\epsilon$, all three auroral 
indices, and the square of the interplanetary magnetic field $B^2$, to obtain the inherent scaling. These studies showed most of the Hurst exponents to be subdiffusive  [7], with H ranging from 0.28 to 0.52. 
This is the most reliable result and it  confirms the fractal nature  
of the  solar wind. Overall, the observed data suggest 
subdiffusive behavior, which is drastically different from the 
superdiffusive behavior demonstrated by the widely used anomalous
diffusion models utilizing the concept of Levy flights [1, 2, 23].  

Attempts at resolving this ambiguity [24, 25] (see also [4, 5])  point out that only some of the parameters reveal the actual scaling 
corresponding to monofractals. A good example is $B^2$ for the solar wind 
parameter [4]. In other cases either the 
generalized structure function (GSF) (depending on its order m) or the 
probability distribution function (PDF), cannot be rescaled. Moreover, 
typically the effective H, inferred from the GSF of the order m=1, is more 
than 0.5, suggesting a superdiffusion, while H inferred for $m>1$ is equal 
to or less than 0.5, suggesting a subdiffusion. Another reason mentioned 
by Watkins et al. [25] is that almost any simulation of fractional Levy motion is effectively one of truncated Levy motion, which may have the $\zeta$  parameter of the GSF superdiffusive corresponding to the case for m=1 and diffusive (H=0.5) for m=2.

Thus, the GSF analysis shows that in many cases the solar wind and 
magnetospheric data do not show ideal scale invariance, that is, 
they do not correspond to a monofractal. And as a result, the simplest 
conclusions in terms of the Hurst exponents are too ambiguous and are not 
clear enough even to distinguish between sub- and superdiffusion behavior. 
On the other hand, Hnat et al. [5] revealed a  collapse of 
the rescaled PDFs for the solar wind parameter $\epsilon$ and the AE index, 
suggesting that some form of simple scale invariance may exist and may 
even allow one to distinguish between these scaling properties for the solar 
wind and the magnetosphere.  According to Hnat et al. 
[22], this does not exclude the possibility that these scaling effects 
can be explained in terms of classical Fokker-Planck diffusion with 
scale-dependent diffusion coefficients.  Finally, a fairly general analysis
of the data [7] increases the level of ambiguity in the interpretation of  the data.
 
The pitfalls of   the data analysis in the
case of multiscale processes, discussed in Section 1, leads to a couple of comments on the results from some recent studies  [7, 8, 22].   First, in these papers the dispersion of the data in the tails is too high   to draw a clear conclusion on whether  a  power-law  tail
  exists. Second, the claim of a possible multifractality is not specified and, in
fact, such behavior has two origins. One is that the data can have a few
different scales (as it will be seen from our analysis in the following) in different intervals of $t$. This may not be a surprise since different physical
processes may be responsible for different intervals of the magnetospheric
data. The second source of multifractality is  related to the so-called
log-periodicity  [1], i.e. a small modulation of the scaling law in $\log t$
scale. This effect, while not discussed in  [7, 8, 22], can be  seen in   Figs~1,~3,~4 of [7]. The existence of
log-periodicity requires very different data analysis [1] in
which the behavior of the higher moments becomes crucial.

The alternative approach to the data analysis and their interpretation, presented in the next section, is aimed at  enhancing the  information
  extracted from the observational data, and at improving the accuracy
of its interpretation. It is worthwhile to add that this approach and technique have been developed   building  on the experience in the study of  different models of dynamical chaos. The study of these models with the new technique has led to the  improvements in its accuracy   and has enabled one  to verify 
some a priori assumptions on the character of processes with flights.\\

\noindent
{\bf 3. ANALYSIS USING L\'{E}VY FLIGHT TECHNIQUE}

Let $y$ be a physical variable of interest, for example,
the density of ions, or a component of the magnetic field,
or the plasma current in a specified direction, and let $y_i$ be its value at discrete time
instant $t_i$. The time instants $\{ t_1 ,\ldots ,t_n \}$ are taken to be
equidistant. It is also assumed that all data $\{ y_i \}$ are collected
at the same location, i.e., the sequence $\{ y_i \}$ is a pure
time-series at a specified point in space. Alternatively, the time series can represent the global behavior, e.g., the indices in the case of the magnetosphere.

The integrated value
$$
S(t)  = \int_0^t dt y (t), \eqno (1)
$$
or, for a discrete time series $\{ t_i \}$,
$$
S_n = S(t_n ) = \sum_{j=1}^n y_j . \eqno (2)
$$
is used in the analysis of flights. In the case when the law of large numbers is valid, the fluctuations of $S_n$
with respect to its mean value $\langle S_n \rangle$ satisfy the
condition
$$
\langle |\delta S(t_N )|^2 \rangle =
  \langle |S_n - \langle S_N \rangle |^2 \rangle \sim N \ , \ \ \ \
N \gg 1 , \eqno (3)
$$
as  in the case of Gaussian processes. In the anomalous case with
power-law tails of the distribution of the values $S_n$, the relation (3)
does not hold since the second or higher moments of the probability distribution
function  of $\Delta S$, $P(\Delta S)$, are infinite with
$$
\Delta S_n \equiv S_n - n (y_n - y_0 ) \eqno (4)
$$
The corresponding processes are of L\'{e}vy type [26, 27]. The new variable $\Delta S_n$ introduced in (4)  differs from the integrated
variable $S_n$ by the subtraction of  a linear trend. This replacement is done for
convenience in working with the data, and it does not influence any physical
interpretation of the results.

\begin{figure}
\centering
\includegraphics[width=15 cm]{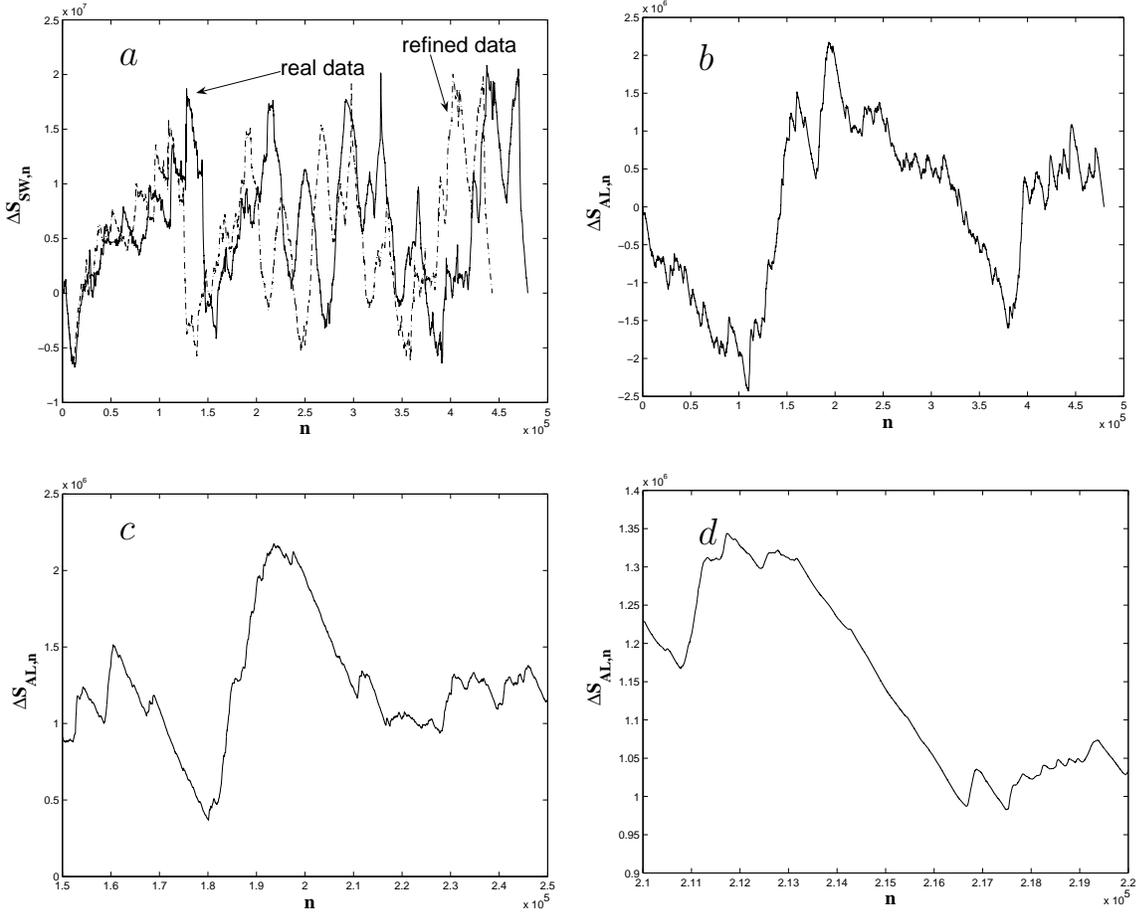}
\caption{\label{fig1} 
Coarse-grained data time series for the deviation $\Delta S_{SW,n}$ (Solar
Wind) and $\Delta S_{AL,n}$ (Auroral Lower index) as a function of discrete
time $n$: 
(a) real data (solid line) for  $\Delta S_{SW,n}$ and refined data  with extrapolation of
the data values for data gaps (dot-dashed line);
(b) real data for  $\Delta S_{AL,n}$;
(c) and (d) - zoom of the corresponding curves in (a) and (b).}
\end{figure}

The solar wind induced electric field and auroral electrojet index [11] are used to compute $\Delta
S_{{\rm SW},n}$ and $\Delta S_{{\rm AL},n}$
and the results are shown in Fig. 1.  The results show large and
small scale wiggles. A few zooms in Fig. 1(c,d) show a kind of
self-similarity
of the behavior of $\Delta S_n$, i.e.,  an indication of the presence of multiscales. Further
analysis of the data can be done using a kind of pattern recognition scheme.
Consider a set of connected values in Fig. 1 as a curve with many different
segments and each segment is defined as an interval of a monotonic behavior
of the curve $\Delta S_n$. Monotonicity of the curve $\Delta S_n$ within
an interval $\Delta n_a = (n^{(1)}, n^{(2)})$ will be defined up to some
interval $a$ which is the interval of coarse-graining. This means that
possible values of $\Delta n_a$ form the set
$$
\Delta n_a \in \{ \Delta n_a \} = \{ a, 2a, 3a, \ldots \} \eqno (5)
$$

The monotonic piece of the curve $\Delta S_n$ within an interval $\Delta n_a$
will be called a ``flight'' and $\Delta n_a$ is the duration of the flight.
All flights are defined up to a smoothing interval $a$. The flights can be
defined in both directions, i.e., as monotonic increases, or decreases
of $\Delta S_n$ within an interval of monotonicity.

The set of data of length $\sim 10^5$  is used to compute the probability distribution function 
$P_{\rm SW} (n_a )$ and $P_{\rm AL} (n_a )$, i.e. the distribution of the
time duration of flights for the solar wind and auroral lower index, respectively. The
corresponding results are shown in Fig. 2 for three different values of
$a = 2, 4 ~and~ 6$. It is evident that the resulting curves do not differ strongly. The results for $n_2$ ($a =2$) are chosen for detailed analysis, without a loss of generality.

\begin{figure}
\centering
\includegraphics[width=15 cm]{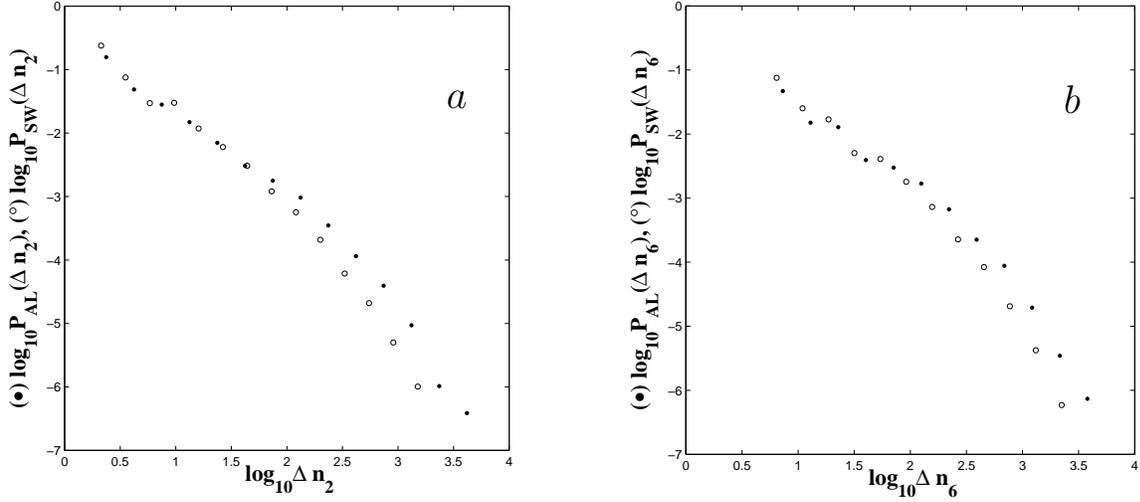}
\caption{\label{fig2} 
Plot of the distributions $P(\Delta n)$ of the time-length $\Delta n$ of
``flights'' for the solar wind (SW) and the magnetosphere (AL), with averaging over $a=2$ (a), and $a=6$ (b) points
along $n$.}
\end{figure}


Both the distributions $P_{\rm SW} (n)$ and $P_{\rm AL} (n)$ have three
characteristic segments defined by intervals of $n$:
zone 1 (0 - 10), zone 2 (10 - 300), zone 3 ($>$ 300). In the first and third
intervals the distributions can be written as
$$
P(n) \sim {\rm const}/n^{1+\beta} \eqno (6)
$$
where the exponent $\beta_1$
depends on the type of data and on the interval.
The computed values of the exponents are  
$$
\beta_{\rm AL}^{(1)} \approx \beta_{\rm SW}^{(1)} \approx 0.9 - 1.0
$$
$$
\beta_{\rm AL}^{(3)} \approx  1.3 - 1.8
$$
$$
\beta_{\rm SW}^{(3)} \approx  1.5 - 2.2 \eqno (7)
$$
The big error bar for zone 3 is due to an ill-defined width of the
transition zone 2. The slope corresponding to AL appears to be smaller than that for SW in zone 3:
$$
\beta_{\rm AL}^{(3)} < \beta_{\rm SW}^{(3)}, \eqno (8)
$$
and the accuracy of the computed values is higher, the larger is the value of $a$ for the selected
intervals of coarse-graining.

The transition zone 2 looks smoother for the solar wind data and one can
assume that this zone corresponds to some {\it resonant type} process whose  origin is in the solar wind. The response of the magnetosphere to the corresponding interval of flights duration makes the transition zone sharper.

Similarity of the behaviors of SW and AL data in zone 1 makes it possible
to predict short flight evolution in the magnetosphere. A similar possibility
to predict, though with a different probability, can be applied to zone 3.

There is a possibility to extract additional information from the data in
Fig. 1. It is the change in the magnitude of $S_n$ during a flight $\Delta n_a$.
Let $s$ be such a change and $P_{\rm SW} (s)$, $P_{\rm AL} (s)$  the probability distribution function
for such changes. One can say that $s$ is a
change of the corresponding magnitude for an intermittent process
during the period of monotonicity.
Following the definitions (1), (2), and (4), during the
monotonic flight between time instants $t_1 ,t_2$ the value
of $s$ is
$$
s_{12} = {\rm const.} (t_2 - t_1 ) \ , \eqno (9)
$$
corresponding to the so-called {\it ballistic flights},  or
$$
s_{12} = {\rm const.} (t_2 -t_1 )^2 \eqno (10)
$$
corresponding to the so-called {\it parabolic flights} with a constant
acceleration.
The corresponding distributions are given in Fig. 3 for two cases of
coarse-graining: $a = 2$ and $a = 4$, which do not show a difference. The
results in Fig. 3 provide interesting observations. There are no 3 zones
as in Fig. 2. The level of changes of $s$ in SW is
 larger by almost an order of magnitude 
than in AL as it is seen from Fig. 1. The same difference is seen for the
distribution functions from Fig. 3 within interval of $s \in $
(10 - 300). After
$s > 1000$ the PDF has the form
$$
P(|s|) \sim {\rm const} /|s|^{1+\alpha_1}, \eqno (11)
y$$
with $\alpha_1$ given by
$$
\alpha_{\rm SW} \approx 0.3 - 0.6,
$$
$$
\alpha_{\rm AL} \approx 0.7 - 1 .\eqno (12)
$$

\begin{figure}
\centering
\includegraphics[width=15 cm]{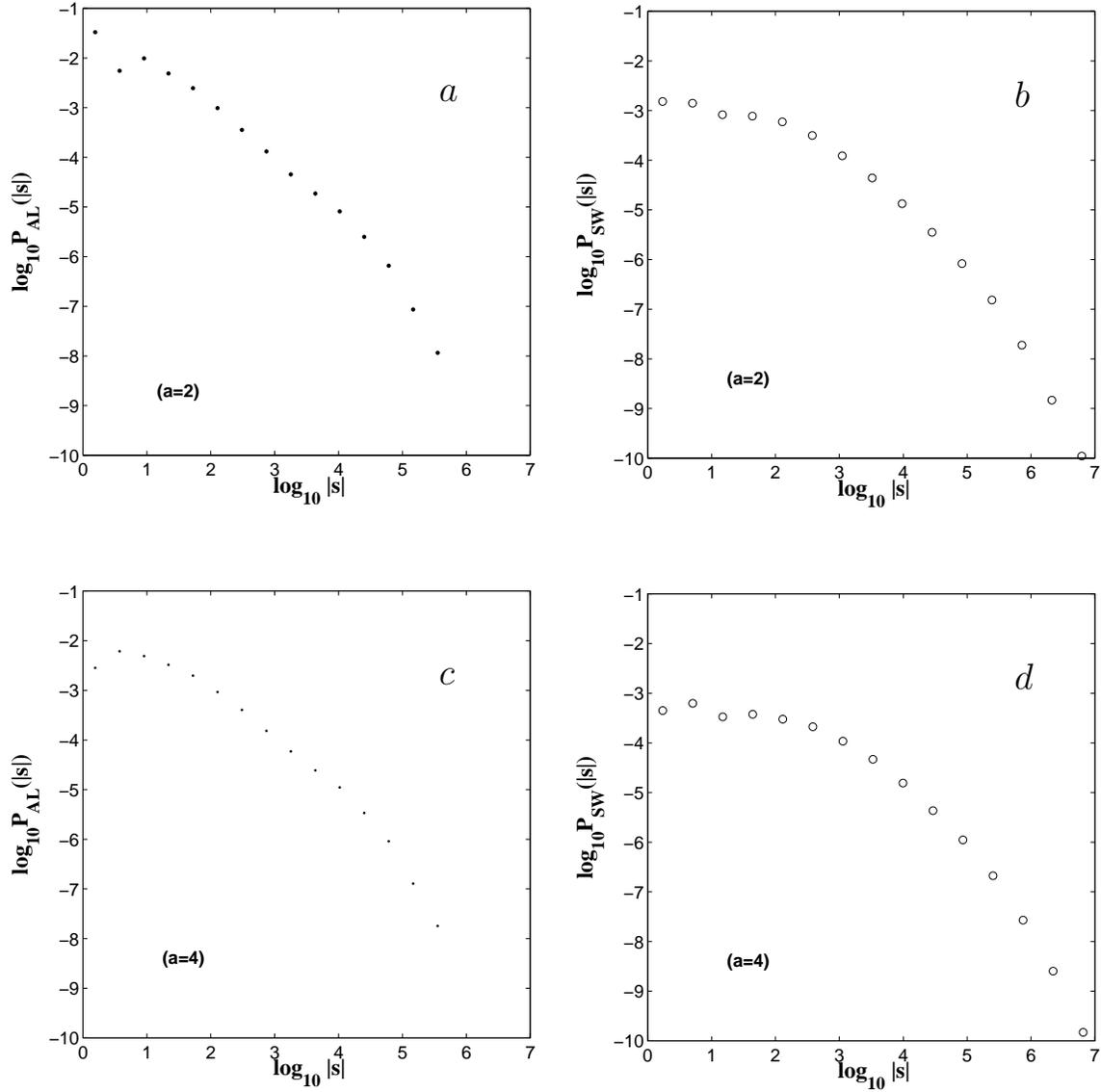}
\caption{\label{fig3} 
Plot of the distributions $P(s)$ of the change $s$ in the magnitude $S_n$ per
flight for the solar wind (SW) and the magnetosphere (AL).}
\end{figure}

The most interesting part of the distributions in Fig. 3 is that for
$s \stackrel{>}{\sim} 1000$, the difference between $P_{\rm AL} (|s|)$
and $P_{\rm SW} (|s|)$ is almost negligible for $s \in$
$(10^3$ - $10^5 ).$ This indicates that in this range of $s$  the magnetosphere and  the solar wind have similar characteristics. The studies of burst life time distributions [28] are closely related to these studies.

 The above values of $\alpha$ correspond to the super diffusive case. This
 is in contrast to the conclusion drawn from the values of the  Hurst
 exponent, which has values $< 0.5$ and thus corresponds 
to the   subdiffusive case.  
As  mentioned in the introduction, the explanation of the Hurst
exponent needs more specific details of the physical processes and the
corresponding theory (Secs. 4, 5; see also discussion in [4, 5, 25]). It should be noted that only some 
of the parameters reveal the actual scaling, corresponding to 
monofractals. A good example is the case of $B^2$ in the solar wind [4]. In other cases either the generalized structure function (depending  on 
its order m) or   the PDF cannot be rescaled, or both. Moreover, 
typically (say, for AE) the effective H, inferred from the $m=1$ GSF,  is more 
than 0.5, suggesting a superdiffusion, while H inferred for $m>1$ is equal 
to or less than 0.5, suggesting a subdiffusion.
It may be noted  
that  a truncated Levy motion may also have the $\zeta$ parameter 
of the GSF corresponding to superdiffusive for m=1 and diffusive ($H_{effective}$=0.5) for 
$m=2$. The inconclusive nature of these studies was discussed in the previous section.\\ 

\noindent
{\bf 4.  MULTISCALE FEATURES IN OBSERVATIONAL DATA AND THEIR MATHEMATICAL MODELS}

Many systems such as the coupled solar wind - magnetosphere system described in section 2 exhibit multiscale behavior in the form of power-law distribution of scale sizes or long tails in the probability distribution functions. The mathematical framework for modeling these phenomena, based on the diffusion-type equations,  are described in this section.

For a system that exhibits self-similarity,  the probability distribution function of a variable 
$y(t)$ can be expressed as
$$
F(y,t)dy = F(y/t^{\nu} ) dy/t^{\nu} = F(\xi ) d\xi \ , \ \ \ \
(\xi = y/t^{\nu} ). \eqno (13)
$$
The exponent $\nu$  is to be determined from the  dependence of the first moment on $t$ as
$$
\langle |y| \rangle = \int |y| F(y,t) dy = {\rm const.} t^{\nu}, \eqno (14)
$$
with the symmetric $F(y,t)$ and the normalization condition
$$
\int F(y,t) dy = 1. \eqno (15)
$$
For example, if $y$ is a coordinate that is subject to a diffusive
type random walk process, then
$$
\mu = 2\nu  \eqno (16)
$$
and $\mu$ is called the transport exponent since for the second moment
$$
\langle y^2 \rangle = {\rm const.} t^{\mu} \ . \eqno (17)
$$

The expression (17) needs further comments. The regular, or normal, diffusion
has $\mu = 1$. Different values of $\mu$, including $\mu = 1$, can result
from the Fokker-Planck type equation
$$
{\partial F(y,t) \over \partial t} =
  {\partial\over\partial y}
  {\cal D} (y)
{\partial F(y,t) \over \partial y}, \eqno (18)
$$
depending on the function ${\cal D} (y)$ that describes nonuniformity of the
diffusion coefficient [29]. The main feature of the diffusion process
is that $F(\xi )$ decays exponentially as $\xi \rightarrow\pm \infty$ and,
as a result, all moments are finite:
$$
\langle y^n \rangle < \infty \ , \ \ \ \ 0 < t < \infty  \eqno (19)
$$
for all $n$.

Another situation, when Eq. (19) is valid is the so-called case of
sub-diffusion [30] described by
$$
{\partial^{\beta} F(y,t) \over \partial t^{\beta} } =
  {\partial\over\partial y}
  {\cal D} (y)
{\partial F(y,t) \over \partial y} \eqno (20)
$$
where
$$
0 < \beta < 1. \eqno (21)
$$
In Eq. (20) the time variation is described in terms of a  fractional derivative of order $\beta$, and this is an example of a fractional kinetic equation. This and other forms of fractional kinetic equations will be discussed in more detail in the following.  For ${\cal D} (y)
= {\cal D}_0 =$ const., Eq. (20) gives
$$
\langle y^2 \rangle = {\rm const.} t^{\beta} , \eqno (22)
$$
which is slower than normal diffusion as a result of Eq. (21). In the both cases of 
Eqs. (18) and (20) $F(y,t)$ have a characteristic width
$$
\Delta\xi = \Delta y /t^{\nu} \eqno (23)
$$
such that for $\xi \gg \Delta\xi$ the values of $F(\xi )$ are
exponentially small, which implies finite moments
$\langle y^n \rangle$
at any finite $t$.

The situation is very different when the probability distribution function $F(y,t)$ behaves as
$$
F(y,t) \sim {c(t) \over y^{1+\delta} } \ , \ \ \ \
\delta > 0 \ , \ \ \ t\rightarrow\infty \ , \ \ \
y\rightarrow\infty ,
\eqno (24)
$$
where $c(t)$ is a function of $t$ alone.   Then exponent $\delta$ can be obtained as the
slope in a $\ln F(y,t)$ vs. $\ln y$ plot. An important part of the data-derived models  is a
plot of a histogram of this dependence. The elementary connection between
variations
$$
\Delta\ln y = (1/|y|) \Delta y \eqno (25)
$$
for $y\rightarrow\infty$ shows that a collection  of the data
of $\ln F(y,t)$ into equal bins
along the $\ln y$ axis would provide less dispersion than
the collection of the data into
equal bins along the $y$ axis. This statement is demonstrated in Fig. 4, where the log of the probabilities $P(\Delta n)$ for $a = 2$ is plotted as a function of $ln \Delta n$ with bins of equal size in $\Delta n$.

\begin{figure}
\centering
\includegraphics[width=15 cm]{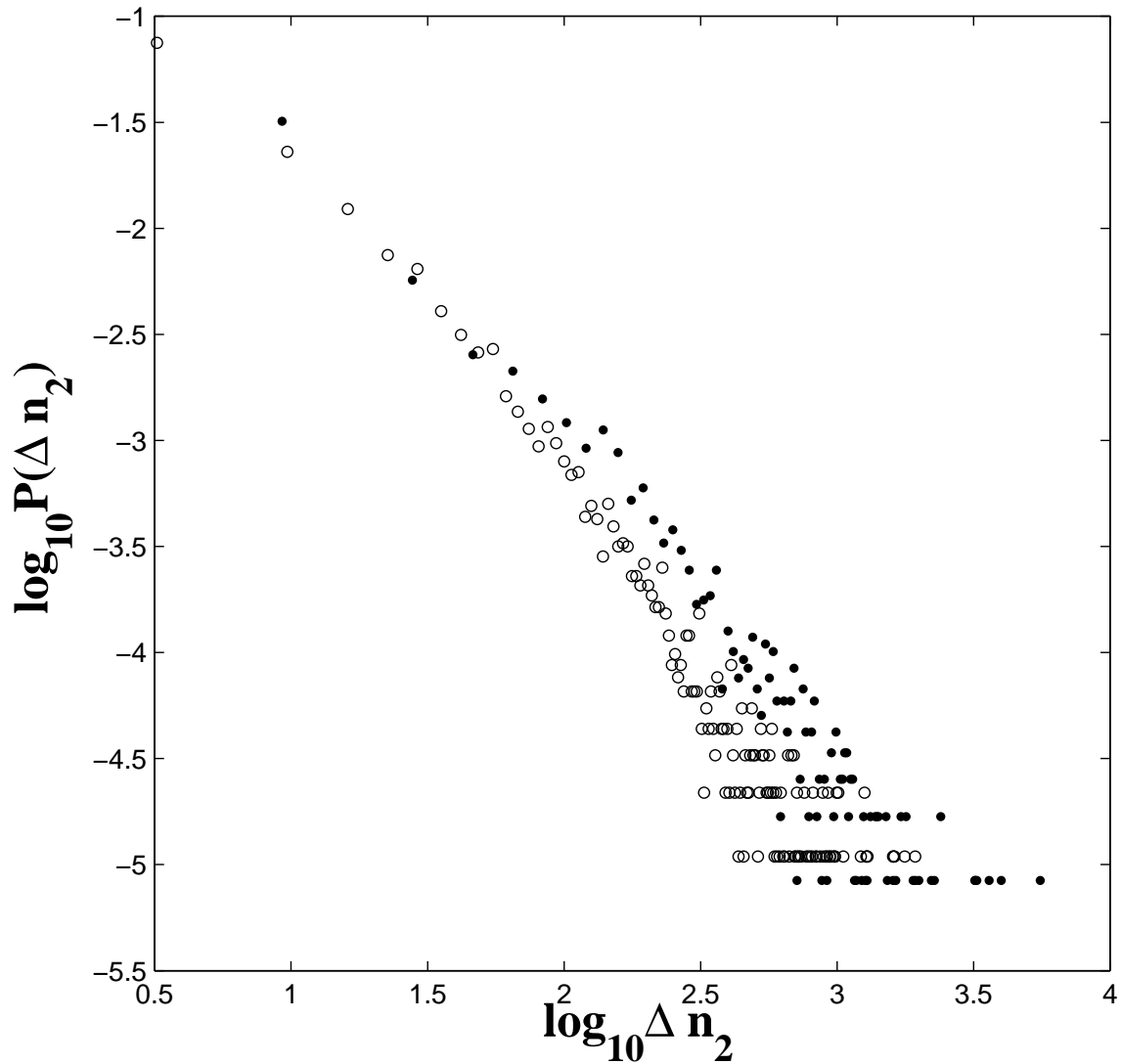}
\caption{\label{fig4} 
Log-Log plot of the distribution function  $P(\Delta n)$ with bins of
equal sizes (equal to 2) in $\Delta n$, as compared  to Fig.~2 where the bin sizes are equal
in $\log \Delta n$. Dots are for the AL data and circles are for the SW
refined data. The plot indicates a strong dispersion of the distribution,
in comparison with  the plot in Fig.~2.}
\end{figure}

Another, equally important comment is related to the truncation of
$F(y,t)$ for fairly large $y$, i.e., one should consider the  data
for $F(y,t)$ and truncate  part of the data for $|y| > y_{\max}$ with
$y_{\max}$ that is not well defined. Let the experimental or simulation
(called ``raw'') data for a fixed $t$ be within the interval
$y \in (y_0 ,y_1 )$ and $|y_1 | \gg |y_0 |$. The power law distribution (24)
imposes large fluctuations of the data for large $|y|$ with a finite (though not
exponentially small) probability. Independent of how big   $t$ is and
how much larger  $y_1$ is compared to $y$,
there will always be a lack of statistics close to
$y_1$, and this is why some data within an interval $(y_{\max} ,|y_1 |)$ has to be excluded.

There is a complementary result for the truncation of the raw data in  all the moments
$$
\langle |y|^m \rangle_{\rm tr} < \infty \ , \ \ \ \ \
(\forall m) ,  \ \ \ m>0 \eqno (26)
$$
are finite and this permits the computation of the self-similarity and even
the multi-fractality parameter $\zeta (m)$ as
$$
\langle |y|^m \rangle_{\rm tr}  \sim t^{\zeta (m)}, \ \ \ \ \
(\forall m) ,  \ \ \ m>0
\eqno (27)
$$
for the function $F(y,t)$ if it exists. In the monofractal case
$$
\zeta (m) = m \mu /2,  \eqno (28)
$$
as compared to Eq. (17).

In reality, the multi-fractality can be due to the so-called
log-periodicity [1], as discussed earlier. In particular, this means that there is some ambiguity in obtaining $\zeta (m)$,
$\mu$, $\delta$, etc., which depend on $y_{\max}$. Thus for any   exponent obtained from data, the interval for which it is computed   and the quality of the  statistics should be specified. Comparison of the exponents from different data sets must be
within similar intervals of the definition of the exponents and for
similar statistics.\\

\noindent
{\bf 5. FRACTIONAL PHENOMENOLOGY AND FRACTIONAL}

{\bf KINETICS}

The crucial feature of the probability distribution function $F(y,t)$ is the existence or
non-existence of the stationary distribution $F(y)$ for $t\rightarrow\infty$.
In many physical problems there is no stationary distribution and the
phase space of the problem has at least one unbounded variable. The plasma density fluctuations in laboratory fusion experiments (tokamaks)  have been modeled in terms of a fractional kinetic equation [31]. Another example is the multiscale nature of the magnetosphere
where the correlated data of the solar wind - magnetosphere system shows clear non-Gaussian PDF [32].   Zaslavsky [23] proposed    a phenomenological  approach to the nonstationary case of the
 evolution of $F(y,t)$ in the presence of multiscale processes.

The basic approach is to write down a balance equation
$$
\delta_t F(y,t) = \overline{\delta_y F(y,t)} ,\eqno (29)
$$
where the bar implies averaging over all admissible paths. This equation can be written as
$$
\delta_t F(y,t) = \bfDelta {_t^{\beta}} F(y,t) \eqno (30)
$$
where $\bfDelta {_t^{\beta}}$ is a generalized difference operator for a time shift by $\Delta t$
parameterized by $\beta$ (see below). Similarly
$$
\delta_y F(y,t) = \sum_{\Delta y} \bfDelta {_y^{\alpha}}
   [ {\cal A} (\Delta y ,y) F(y,t)] \eqno (31)
$$
where $ \bfDelta {_y^{\alpha}}$ is a generalized difference operator for a $y$-variable shift by
$\Delta y$ along a path ${\cal A} (\Delta y,y)$ in the phase space,
parameterized by $\alpha$, and the line in (29) implies averaging over all
admissible paths. Definitions of $\bfDelta_t^{\beta}$ and
$\bfDelta_y^{\alpha}$ are given in [1] and their main properties are
$$
\bfDelta {_t^{\beta}} \sim (\Delta t)^{\beta}
  {\partial^{\beta} \over \partial t^{\beta}}  \ , \ \ \ \
(\Delta t \rightarrow 0)
$$
$$
\bfDelta {_y^{\alpha}} \sim (|\Delta y|)^{\alpha}
  {\partial^{\alpha} \over \partial |y|^{\alpha} } \ , \ \ \ \
(|\Delta y| \rightarrow 0) \eqno (32)
$$
where   the fractional derivatives of the orders $\beta$ and
$\alpha$, respectively, are introduced. These derivatives work in
different ways since there are different definitions of the variables $t\in (0,\infty )$ and $y\in (-\infty ,\infty )$.

To simplify the problem we consider $F(y,t)$ to be symmetric with respect to $y$
and ${\cal A} = {\cal A} (\Delta y)$. Then the balance equation can be
written in the form [zaslavsky02, zaslavsky00; 1, 23]:
$$
{\partial^{\beta} F(y,t) \over \partial t^{\beta} } =
{\cal D}_{\alpha\beta}
{\partial^{\alpha} F(y,t) \over \partial |y|^{\alpha} }
\eqno (33)
$$
where
$$
{\cal D}_{\alpha\beta} =
  \overline{ {|\Delta y|^{\alpha} \over (\Delta t)^{\beta} } {\cal A}
  (\Delta y) } \ , \ \ \ \
  {\cal A} (0) = {\rm const.} \eqno (34)
$$
Eq. (33) is a fractional kinetic equation (FKE) [1, 23] since
$(\alpha ,\beta )$ can be fractional. In the case of $\beta = 1$,
$\alpha = 2$, Eq. (33) is a regular diffusion equation with the diffusion
coefficient ${\cal D}_{\alpha\beta}$. The important part of the derivation
of (33) is the existence of finite ${\cal D}_{\alpha\beta}$ in the limit
$$
\Delta t \rightarrow 0 \ ,  \ \ \ \
\Delta y \rightarrow 0 \ ,  \ \ \ \
{\cal D}_{\alpha\beta} = {\rm const.} \ , \eqno (35)
$$
known, for the case $\beta = 1$, $\alpha = 2$, as the Kolmogorov condition
[1]. The existence of the limit (35) is a nontrivial fact, and eventually
it defines the variable $y$ for which the FKE can be written.

An important property of Eq. (33) and the definition (34) is the rescaling
invariance. Consider the following rescaling
$$
\Delta y \rightarrow \lambda_y \Delta y \ , \ \ \ \
\Delta t \rightarrow \lambda_t \Delta t \eqno (36)
$$
that gives
$$
{\cal D}_{\alpha\beta} \rightarrow
  (\lambda_y^{\alpha} /\lambda_t^{\beta} )
{\cal D}_{\alpha\beta} \ . \eqno (37)
$$
The FKE Eq. (33) is scale-invariant under the transforms (36) and (37) if
$$
\beta / \alpha = \ln\lambda_y /\ln\lambda_t \ . \eqno (38)
$$
On the other hand,   multiplying Eq. (33) by $|y|^{\alpha}$ and integrating both
sides over $y$, yields
$$
\langle |y|^{\alpha} \rangle_{\rm tr}  =
{\rm const.} \ t^{\beta} \ , \eqno (39)
$$
i.e. the transport exponent
$$
\mu = 2\beta /\alpha , \eqno (40)
$$
or using (38)
$$
\mu = 2\ln\lambda_y /\ln\lambda_t \ . \eqno (41)
$$

For $\beta = 1$, $0 < \alpha < 2$, the FKE describes the L\'{e}vy process.
The moments
$\langle |y|^p \rangle$
are finite if
$p < \alpha$ [30] and Eq. (39) can be considered only for the
truncated $F(y,t)$ as was commented on earlier.

In simple terms, if the data of a physical system exhibit the scaling  property given by Eq. (36),
then the FKE, Eq. (33), can be considered as a model of the   processes
the data represent.\\  

\noindent
{\bf 6. FKE FROM THE DATA}

The modeling of a physical system, whose data yields
  the scaling properties described in
Sec. 2, in terms of a fractional kinetic equation is discussed in this section. There exists an indefiniteness in the reconstruction of the  parameters $(\alpha ,\beta )$  that define the FKE. Starting with the rescaling parameters ($\lambda_y ,\lambda_t$) obtained from data,  only the ratio $\beta /\alpha$, but not the independent values of $(\alpha ,\beta )$ can be obtained. However, this indefiniteness does not influence the transport exponent $\mu$. Another
indefiniteness of the data analysis arises from the Kolmogorov condition (35)
since  this information can not be obtained from the  the time series data alone,
and other approaches need to be employed.  

In  Sec. 2 the data of the solar wind - magnetosphere system  were analyzed to obtain   the distributions
$P(s)$ and $p(t)$. Since $s$ is the accumulated variations in the variable, for example, the magnetic field in the solar wind
during the time $t$ of the duration of a flight, it is worthwhile
to consider $y = s/t$ as a measure of the physical variable  within a flight. From the same data, one can obtain distributions presented
in Fig. 5. By choosing
$$
P(s/t) = P(y) = {{\rm const.} \over y^{1+\alpha_2} } ,\eqno (42)
$$

\begin{figure}
\centering
\includegraphics[width=15 cm]{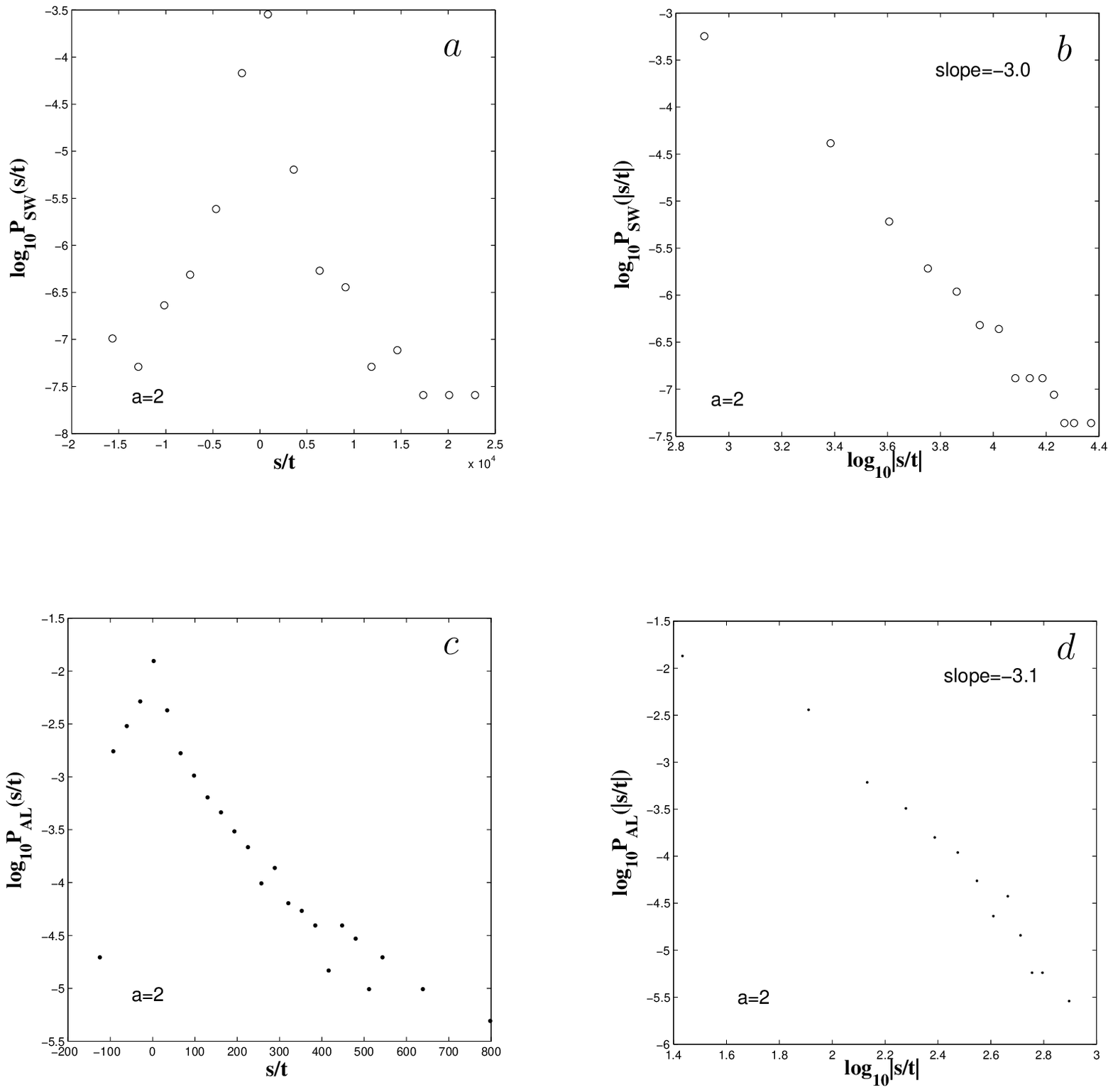}
\caption{\label{fig5} 
Distributions $P(s/t)$ for SW (a) and AL (c) and their
tails (b) and (d).
}
\end{figure}

where   $\alpha_2$ can be computed in the cases of the solar wind and the auroral electrojet index, and this yields
$$
\alpha_{\rm SW} \approx 2.0 \ , \ \ \ \ \
\alpha_{\rm AL} \approx 2.1 \ . \eqno (43)
$$
Then the 
results (7) and (43) give
$$
\beta_{\rm SW}^{(1)} /\alpha_{\rm SW} \approx
  \beta_{\rm AL}^{(1)} /\alpha_{\rm AL} \approx 0.45 - 0.5
$$
$$
\beta_{\rm SW}^{(3)} /\alpha_{\rm SW} \approx 0.75 - 1.1
$$
$$
\beta_{\rm AL}^{(3)} /\alpha_{\rm AL} \approx 0.62 - 0.86
\eqno (44)
$$
The results (44) show that processes SW and AL are very close to each
other in zone 1, related to small scale flights, and fairly different in
zone 3 that is related to large scale flights.

In addition, for short flights the corresponding value of $\mu$ is close to
the normal one and even could be related to subdiffusion $(\mu < 1)$ while
for the long flights $\mu > 1$ and is related to superdiffusion processes
if the variable $y = s/t$ is the right one with respect to the Kolmogorov
condition.\\

\noindent
{\bf 7. NUMERICAL STUDIES OF FRACTIONAL KINETIC EQUATIONS}
 
In this section numerical results from the investigation of two classes of kinetic equations discussed earlier are presented. The first is the fractional kinetic equation (FKE) as defined earlier, Eq. (33), with $\beta=1$ and $\alpha=1.6$. The second is the Fokker-Planck type equation, Eq. (18), but  with a specific power-law dependence for the diffusion coefficient.  For the sake of definiteness, the two equations are 

\renewcommand{\theequation}{45}
\begin{equation}
\frac{\partial F(y,t)}{\partial t}=\frac{\partial^{\alpha} F(y,t)}{\partial |y|^{\alpha}}
\end{equation} 

\renewcommand{\theequation}{46}
\begin{equation}
\frac{\partial F(y,t)}{\partial t}=\frac{\partial}{\partial y}[ |y|^{\mu} \frac{\partial F(y,t)}{\partial y}].
\end{equation}

For these studies $\mu=2-\alpha$ to make the mean-squared displacement scale similarly with time for the two equations. For the FKE, a specific choice of the fractional derivative in y has been made. For this choice, the Fourier transform representation for the dependent variable yields      

\renewcommand{\theequation}{47}
\begin{equation}
\int_{-\infty}^{ \infty} dy e^{iky}\frac{\partial^{\alpha} F(y,t)}{\partial |y|^{\alpha}}=-|k|^{\alpha}{\hat F}(k,t)
\end{equation}
Here ${\hat F}(k,t)$ is the Fourier transform of $F(y,t)$. Thus the solution in Fourier space of Eq. (45) is
\renewcommand{\theequation}{48}
\begin{equation}
{\hat F}(k,t)= {\hat F}(k,t_0)e^{-|k|^{\alpha}(t-t_0)}
\end{equation}
For Eq. (46) an implicit finite difference scheme yields a tridiagonal
equation which can be readily inverted to obtain the distribution function
$F(y,t)$ 
at different instants of time. Although the scheme is a stable one, the time-stepping restriction is determined by the requirement on accuracy.

\begin{figure}
\centering
\includegraphics[width=10 cm]{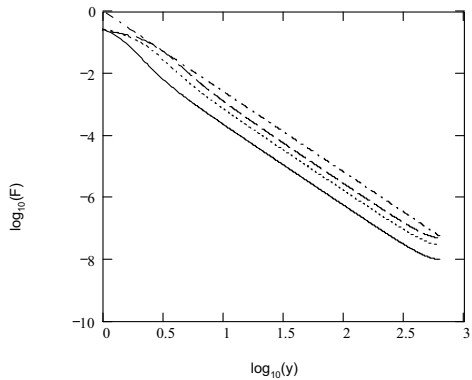}
\caption{\label{fig6} 
$log_{10}(F)$ versus $\log_{10}(y)$ for $t=0.5$ (solid), $t=1.0$ (dot), $t=1.5$ (dash), asymptotic theory (dot-dash).}
\end{figure}

Fig. 6 is a log-log plot for the solutions of Eq. (45) for $\alpha$=1.6 at
different instants of time. The different lines are for three instants of
time namely,
$t=0.5$ (solid), $t=1.0$ (dot), $t=1.5$ (dash). The last line (dot-dash) is the theoretically determined asymptotic solution derived in Sec. 4 (Eq. 40).

\begin{figure}
\centering
\includegraphics[width=10 cm]{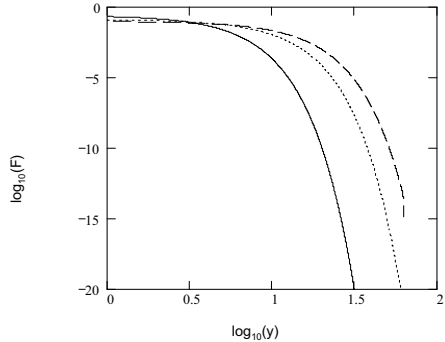}
\caption{\label{fig7} 
$\log_{10}(F)$ versus $\log_{10}(y)$ for $t=2.0$ (solid), $t=6.0$ (dot), $t=10.0$ (dash).}
\end{figure}

In Fig. 7 is shown the log-log plot of the solutions of Eq. 46 for
$\mu$=0.4 also at different instants of time.  Once again the different
lines are at 
$t=2$ (solid), $t=6$ (dot) and $t=10$ (dash). 
For this case the distribution does
not display any power-law type behavior.

\begin{figure}
\centering
\includegraphics[width=15 cm]{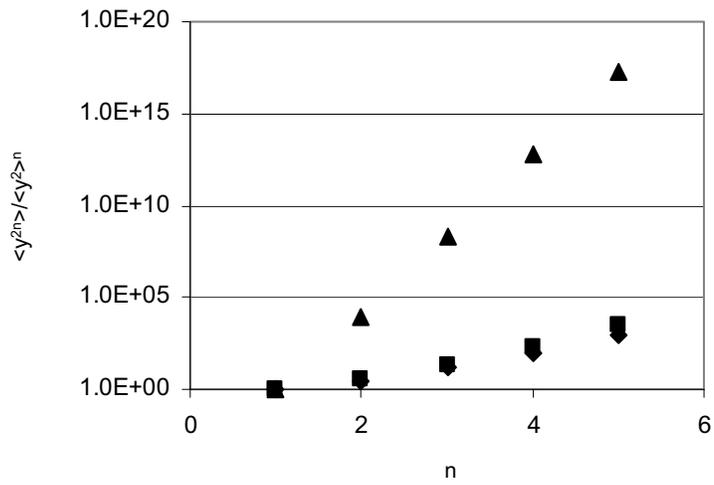}
\caption{\label{fig8} 
$<y^{2n}>/<y^2>^n$ versus $n$ for Gaussian distribution function( diamonds), distribution function obtained from Fokker-Planck like equation (Eq. (46)) (squares) and distribution function obtained from the fractional kinetic equation (Eq. (45)) (triangles).}
\end{figure}
  
Furthermore in Fig. 8, the normalized even moments (to $2n=10$) are plotted
for the Gaussian distribution (diamonds), Eq (46) (at $t=8$) (squares) and
for the FKE at $t=1.5$ (circles). 
Here again it is clear that for Eq. (46) the moments are finite and comparable to the Gaussian case. 
However for the FKE the moments are large and dependent on the size of the
computation box (i.e. maximum value of y used in the computation) 
and in principle are infinite as expected, based on the arguments presented 
in Sec. 4.\\      
 
\noindent
{\bf 8. CONCLUSION }

 In the solar wind-magnetosphere interaction, multiscale
features coexist
along with the global or coherent features, and have been studied
extensively using nonlinear dynamical techniques. The detailed
properties of these
phenomena are studied using Levy flights and
fractional kinetics models.  In the  solar wind - magnetosphere coupling, a technique to analyze the Levy-type processes is applied to the time series data of the solar wind electric field and the auroral
electrojet index. The probability distribution function of the flights
show similarities and differences, and provides a new insight into the
origin of the multiscale behavior through the
different values of the scaling indices. In a complementary approach,
the fractional kinetic equations, which uses   fractional derivatives
to represent the complexity,  provide a suitable mathematical
framework for the multiscale behavior. Unlike the usual diffusion
equations, the solutions of these equations yield non-convergent
moments, showing its multiscale features.  The origin of multiscale
behavior is studied by using a fractional kinetic equation and a
diffusion equation with a space dependent diffusion coefficient.
Numerical solutions of these equations are used to analyze the nature
of the equations by computing the moments. The fractional kinetic
equations yield solutions with large moments and are divergent.
On the other hand the solutions of the diffusion equation have
convergent moments similar to those of Gaussian distributions. These
results lead to the conclusion that the fractional kinetic equations
are suitable models of multiscale phenomena.

In the the study of the solar wind by Hnat et al [5], all the
probability distribution functions, e. g., their Figs. 3, 4, and 5, have
infinite first moment because of the small Hurst exponent.   However this
does not follow from their equations (Eqs.(2) and (3)).   This
inconsistency raises the question of what leads to multiscale properties.
In the studies presented in this paper, only the fractional kinetic
equations lead to the  formation of tails in the distribution functions, 
and thus   can be considered as   appropriate models of
multiscale processes. It should however be noted that the parameters defining  fractional kinetic equation could be improved as more data become available and with a better understanding of the magnetosphere dynamics.\\  

{\bf ACKNOWLEDGMENTS. }

This work was supported by NSF / CMG grant DMS-0417800, and ONR grant N00014-02-1-0056. The simulations were supported by NERSC. The correlated data set of the solar wind - magnetosphere system used in this paper was compiled by Jian Chen.

\begin{center}
{\bf REFERENCES}
\end{center}

\noindent
[1] G.M. Zaslavsky, Phys. Reports 371, 461 (2002)

\noindent
[2] G.M. Zaslavsky et al.,  Phys. Plasmas, 7,  3691 (2000).

\noindent
[3] J. Takalo and J. Timmonen, Geophys. Res. Lett., 21, 617, 1994. 

\noindent
[4] B. Hnat et al., Geophys. Res. Lett., 29 (10), 10.1029/2001GL014587, 2002.

\noindent
[5] B. Hnat et al., Geophys. Res. Lett., 30, 2174, 2003.  

\noindent
[6] S. C. Chapman et al., Submitted to Nonl. Proc. in Geophys., (2005).

\noindent
[7] N. W. Watkins et al., arXiv Physics/0509058, v. 1, (2005).

\noindent
[8] L. Gil and D. Sornette,  Phys. Rev. Lett., 76, 3991 (1996).

\noindent
[9] H. E. Hurst,  Trans. Am. Soc. Civ. Eng., 116, 770 (2005).

\noindent
[10] J. Feder,  ``Fractals'' (Plenum Press, New York, 1988).

\noindent
[11] J. Chen, A. S. Sharma,   J. Geophys. Res., in press, 2005.  

\noindent
[12] M. F. Shlesinger M.F. and M. A. Coplan, J. Stat. Phys. 52, 1423 (1988).

\noindent
[13] L. F. Burlaga, "Interplanetary Magnetohydrodynamics", (Oxford Univ. Press, 1995).

\noindent
[14] G. Consolini et al., Phys. Rev. Lett., 76, 4082 (1996). 

\noindent
[15] B. T. Tsurutani et al., Geophys. Res. Lett., 17, 279 (1990).

\noindent
[16] G. Consolini, in "Cosmic Physics in the Year 2000", (1997). 

\noindent
[17] P. N. Mayaud,  ``Derivation, Meaning, and Use of Geomagnetic Indices'' (Amer. Geophys. Union, Washington, DC, 1980).

\noindent
[18] D. Vassiliadis et al., J. Geophys. Res., 100, 3495 (1995).

\noindent
[19] A. Y. Ukhorskiy et al., Geophys. Res. Lett., 31, L08802 (2004).

\noindent
[20] V. M. Uritsky and M. I. Pudovkin, Ann. Geophys., 16, 1580 (1998).  

\noindent
[21] C. P. Price and D. E. Newmann, J. Atm. Sol.Terr. Phys., 63, 1387 (2001).   

\noindent
[22] B. Hnat et al., Phys. Rev. E, 67, 056404 (2003). 

\noindent
[23] G.M. Zaslavsky, Physica D, 76, 110 (1994);
Chaos, 4, 25 (1994).

\noindent
[24] N. W. Watkins et al., Fractals, 9, 471 (2001).
 
\noindent
[25] N. W. Watkins et al., Space Sci. rev., in press (2005).

\noindent
[26] P. L\'{e}vy, ``Theorie de L'Addition Variables Aleatoires'' (Guthier-Villars, Paris, 1937).

\noindent
[27] V.V. Uchaikin and V.M. Zolotarev,
``Chance and Stability'', (VCP, Utrecht, 1999).
 
\noindent
[28] M. P. Freeman et al.,  Geophys. Res. Lett., 27, 1087 (2000).
   
\noindent
[29] H. Risken,
``The Fokker-Planck Equation: Methods of Solution and Applications'',
(Springer, Berlin, 1996).

\noindent
[30] A.I. Saichev and G.M. Zaslavsky,
Chaos {\bf 7}, 753.
 
\noindent
[31] B. A. Carreras et al.,   Phys. Plasmas, 8, 5096 (2001). 

\noindent
[32] A. Y. Ukhorskiy et al.,   J. Geophys. Res., 107, 1369 (2002).

\end{document}